\documentclass[aps,pra,superscriptaddress,preprint,onecolumn,%
longbibliography]{revtex4-1}
\usepackage{amsfonts}
\usepackage{amsmath}
\usepackage{amssymb}
\usepackage{amsthm}

\DeclareMathOperator{\codim}{codim}
\newcommand{\Eig}[1]{\textrm{Eig}(#1)}
\newcommand{\ascangle}[2]{\theta^\uparrow_{#2}(#1)}
\newcommand{\desangle}[2]{\theta^\downarrow_{#2}(#1)}
\newcommand{\desabsangle}[2]{|\theta|^\downarrow_{#2}(#1)}
\newcommand{\desket}[2]{|\phi^\downarrow_{#2}(#1)\rangle}
\newcommand{\desbra}[2]{\langle\phi^\downarrow_{#2}(#1)|}
\newcommand{\desspace}[2]{{\mathcal H}^\downarrow_{#2}(#1)}
\newcommand{\desallspace}[1]{{\mathcal H}^\bullet(#1)}

\newtheorem{definition}{Definition}
\newtheorem{theorem}{Theorem}
\newtheorem{lemma}{Lemma}

\begin{document}

\title{Elementary Proofs Of Two Theorems Involving Arguments Of Eigenvalues Of
 A Product Of Two Unitary Matrices}

\author{H.~F. Chau}\email{hfchau@hku.hk}
\affiliation{Department of Physics, University of Hong Kong, Pokfulam Road,
 Hong Kong}
\affiliation{Center of Computational and Theoretical Physics, University of
 Hong Kong, Pokfulam Road, Hong Kong}
\author{Y.~T. Lam}\thanks{Present address: Department of Mathematics,
 University of Hong Kong}
\affiliation{Department of Physics, University of Hong Kong, Pokfulam Road,
 Hong Kong}
\date{\today}

\begin{abstract}
 We give elementary proofs of two theorems concerning bounds on the maximum
 argument of the eigenvalues of a product of two unitary matrices --- one by
 Childs \emph{et al.} [J. Mod. Phys., \textbf{47}, 155 (2000)] and the other
 one by Chau [arXiv:1006.3614].  Our proofs have the advantages that the
 necessary and sufficient conditions for equalities are apparent and that they
 can be readily generalized to the case of infinite-dimensional unitary
 operators.
\end{abstract}


\maketitle

 Let $\Eig{U}$ denotes the set of all eigenvalues of a unitary matrix $U$.
 Interestingly, one can give non-trivial information on $\Eig{U V}$, usually in
 the form of inequalities, solely based on $\Eig{U}$ and $\Eig{V}$.  (See, for
 example, Refs.~\cite{eigenvalue_book,matrix_analysis_book} for comprehensive
 reviews of the field of spectral variation theory of matrices, including
 Hermitian and normal ones.)  In this paper, we give elementary proofs of two
 such inequalities.  Let us begin by introducing a few notations first.

\begin{definition}
 Let $U$ be a $n$-dimensional unitary matrix.  Generalizing the conventions
 adopted in Ref.~\cite{matrix_analysis_book}, we denote the arguments (all
 arguments in this paper are in principal values) of the eigenvalues of $U$
 arranged in descending and ascending orders by $\desangle{U}{j}$'s and
 $\ascangle{U}{j}$'s respectively, where the index $j$ runs from $1$ to $n$.
 That is to say, $U = \sum_j e^{i \desangle{U}{j}} \desket{U}{j}\desbra{U}{j}$
 where $\desangle{U}{j} \in (-\pi,\pi]$ and $\desket{U}{j}$ is a normalized
 eigenvector of $U$ with eigenvalue $e^{i \desangle{U}{j}}$.  Moreover, we
 write the eigenspace spanned by the eigenket $\desket{U}{j}$ by
 $\desspace{U}{j}$, and the eigenspace corresponding to the eigenvalue $e^{i
 \desangle{U}{1}}$ by $\desallspace{U}$, respectively.  (Clearly,
 $\desallspace{U} = \desspace{U}{1}$ if and only if $e^{i \desangle{U}{1}}$ is
 a non-degenerate eigenvalue.)  We further denote the absolute value of the
 argument of the eigenvalues of $U$ arranged in descending order by
 $\desabsangle{U}{j}$'s, where the index $j$ runs from $1$ to $n$.
 \label{Def:form}
\end{definition}

 Recently, Childs \emph{et al.}~\cite{principal_bound} proved the validity of
 the following theorem using Baker-Campbell-Hausdorff formula and eigenvalue
 perturbation theory.

\begin{theorem}
 Let $U, V$ be two $n$-dimensional unitary matrices satisfying
 $\desangle{U}{1} + \desangle{V}{1} \leq \pi$ and $\ascangle{U}{1} +
 \ascangle{V}{1} > -\pi$.  Then
 \begin{subequations}
 \begin{equation}
  \desangle{U V}{1} \leq \desangle{U}{1} + \desangle{V}{1}
  \label{E:UV_des_inequality}
 \end{equation}
 and
 \begin{equation}
  \ascangle{U V}{1} \geq \ascangle{U}{1} + \ascangle{V}{1} .
  \label{E:UV_asc_inequality}
 \end{equation}
 \end{subequations}
 Furthermore, the equality of Eq.~(\ref{E:UV_des_inequality}) holds if and only
 if $\dim \left[ \desallspace{U} \cap \desallspace{V} \right] \geq 1$.
 Similarly, the equality of Eq.~(\ref{E:UV_asc_inequality}) holds if and only
 if $\dim \left[ \desallspace{U^{-1}} \cap \desallspace{V^{-1}} \right] \geq
 1$.
 \label{Thrm:desangle_bound}
\end{theorem}

 Actually, a more general version of Theorem~\ref{Thrm:desangle_bound} was
 first proven by Nudel'man and \u{S}varcman~\cite{convex_hull_bound} by looking
 into the geometric properties of certain hyperplanes related to the argument
 of the eigenvalues of a unitary matrix.  Built on this geometric approach,
 Thompson~\cite{direct_extension} extended Nudel'man and \u{S}varcman's result
 by giving an even more general version of Theorem~\ref{Thrm:desangle_bound}.
 (Note that Nidel'man and \u{S}varcman as well as Thompson used a different
 convention in which all arguments of the eigenvalues are taken from the
 interval $[0,2\pi)$.  Nonetheless, the convention does not affect the
 conclusions of Theorem~\ref{Thrm:desangle_bound}.)  Later on, Agnihotri and
 Woodward~\cite{q_calculus} as well as Biswas~\cite{Biswas99} showed among
 other things the validity of Theorem~\ref{Thrm:desangle_bound} by means of
 quantum Schubert calculus.  Belkale~\cite{Belkale01} obtained
 Theorem~\ref{Thrm:desangle_bound} by studying the local monodromy of certain
 geometrical objects.

 Along a similar line of investigation, Chau~\cite{metric} recently showed
 among other things the following theorem using Rayleigh-Schr\"{o}dinger
 series.

\begin{theorem}
 Let $U, V$ be two $n$-dimensional unitary matrices.  Then
 \begin{equation}
  \desabsangle{U V}{1} \leq \desabsangle{U}{1} + \desabsangle{V}{1} .
  \label{E:UV_desabs_inequality}
 \end{equation}
 Moreover, the equality holds if and only if
 \begin{enumerate}
  \item $\desabsangle{U}{1} + \desabsangle{V}{1} \leq \pi$; and
  \item
   \begin{enumerate}
    \item $\dim \left[ \desallspace{U} \cap \desallspace{V} \right] \geq 1$,
     $\desangle{U}{1} = \desabsangle{U}{1}$ and  $\desangle{V}{1} =
     \desabsangle{V}{1}$; or
    \item $\dim \left[ \desallspace{U^{-1}} \cap \desallspace{V^{-1}} \right]
     \geq 1$, $\ascangle{U}{1} = -\desabsangle{U}{1}$ and $\ascangle{V}{1} =
     - \desabsangle{V}{1}$.
   \end{enumerate}
  \end{enumerate}
 \label{Thrm:desabsangle_bound}
\end{theorem}

 Note that all existing proofs of Theorems~\ref{Thrm:desangle_bound}
 and~\ref{Thrm:desabsangle_bound} involve rather high level geometrical or
 analytical methods.  Here, we report elementary proofs of these two theorems.
 One of the advantages of these elementary proofs is that one can easily deduce
 the necessary and sufficient conditions for equalities.  Besides, it is
 straightforward to extend the theorem to cover the case of
 infinite-dimensional unitary operators.

 Our elementary proofs of these two theorems rely on Lemma~\ref{Lemma:min_max}
 which in turn follows from Lemma~\ref{Lemma:product_phase}.

\begin{lemma}
 Let $U, V$ be two $n$-dimensional unitary matrices with $\desangle{U}{1} -
 \ascangle{U}{1}, \desangle{V}{1} - \ascangle{V}{1}, \desangle{U}{1} +
 \desangle{V}{1}, -\ascangle{U}{1} -\ascangle{V}{1} < \pi$.  Then,
 \begin{equation}
  \arg \desbra{U V}{j} U \desket{U V}{j} + \arg \desbra{U V}{j} V
  \desket{U V}{j} = \desangle{U V}{j}
  \label{E:des_product_phase}
 \end{equation}
 for $j = 1,2,\ldots , n$.
 \label{Lemma:product_phase}
\end{lemma}
\begin{proof}
 By definition, $U V \desket{U V}{j} = e^{i\desangle{U V}{j}} \desket{U V}{j}$.
 Since $U$ is unitary, we know that $\desbra{U V}{j} V \desket{U V}{j} =
 e^{i\desangle{U V}{j}}$ $\desbra{U V}{j} U^{-1} \desket{U V}{j} =
 e^{i\desangle{U V}{j}} \left[ \desbra{U V}{j} U \desket{U V}{j} \right]^{*}$.
 By taking the arguments in both sides, we obtain
 \begin{equation}
  \arg \desbra{U V}{j} V \desket{U V}{j} = \desangle{U V}{j} - \arg
  \desbra{U V}{j} U \desket{U V}{j} \bmod 2\pi .
  \label{E:pre_des_product_phase}
 \end{equation}
 Note that for any normalized state ket $|\psi\rangle$, $\langle\psi | U |
 \psi\rangle$ and $\langle\psi | V | \psi\rangle$ are located in the convex
 hull formed by the vertices $\{ e^{i \desangle{U}{k}} \}_{k=1}^n$ and $\{ e^{i
 \desangle{V}{k}} \}_{k=1}^n$ on the complex plane ${\mathbb C}$, respectively.
 Combined with the conditions that $\desangle{U}{1} - \ascangle{U}{1},
 \desangle{V}{1} - \ascangle{V}{1} < \pi$, we have $\arg \desbra{U V}{j} U
 \desket{U V}{j} \in [\ascangle{U}{1},\desangle{U}{1}]$ and $\arg
 \desbra{U V}{j} V \desket{U V}{j} \in [\ascangle{V}{1},\desangle{V}{1}]$.
 Since $\desangle{U}{1} + \desangle{V}{1}, -\ascangle{U}{1} -\ascangle{V}{1} <
 \pi$, we conclude that Eq.~(\ref{E:pre_des_product_phase}) is valid even if
 the modulo $2\pi$ is removed.
\end{proof}

\begin{lemma}
 Let $U$ be a $n$-dimensional unitary matrix with $\desangle{U}{1} -
 \ascangle{U}{1} < \pi$.  Then, for $j = 1,2,\ldots, n$, we have
 \begin{equation}
  \desangle{U}{j} = \min_{{\mathcal H} \colon \codim {\mathcal H} = j-1} \quad
  \max_{|\psi\rangle \in {\mathcal H}} \arg \langle \psi | U | \psi \rangle .
  \label{E:min_max_des}
 \end{equation}
 Furthermore, the extremum in the R.H.S. of the above equation is attained by
 choosing ${\mathcal H} = \bigoplus_{k=j}^n \desspace{U}{k}$.  In particular,
 \begin{equation}
  \ascangle{U}{1} \leq \arg \langle \psi | U | \psi \rangle \leq
  \desangle{U}{1}
  \label{E:min_max_des_special}
 \end{equation}
 for all $|\psi\rangle$.
 \label{Lemma:min_max}
\end{lemma}
\begin{proof}
 Any Hilbert subspace of codimension $j-1$ must have non-trivial intersection
 with the $j$-dimensional Hilbert space $\bigoplus_{k=1}^j \desspace{U}{k}$.
 In addition, the set $S = \{ \langle \psi | U | \psi \rangle : | \psi \rangle
 \in \bigoplus_{k=1}^j \desspace{U}{k} \textrm{~and~} \langle \psi | \psi
 \rangle = 1 \}$ is equal to the convex hull formed by the vertices $\{ e^{i
 \desangle{U}{k}} \}_{k=1}^j$ on the complex plane ${\mathbb C}$.  Since
 $\desangle{U}{1} - \desangle{U}{j} \leq \desangle{U}{1} - \ascangle{U}{1} <
 \pi$, $S$ lies on a half plane on ${\mathbb C}$ and $S$ does not intersect
 with the negative real half-line.  Hence, every normalized vector
 $|\psi\rangle$ in $\bigoplus_{k=1}^j \desspace{U}{k}$ must obey $\arg \langle
 \psi | U | \psi \rangle \geq \desangle{U}{j}$; and the equality holds if
 $|\psi\rangle = \desket{U}{j}$ up to a phase.  (This condition for equality is
 both necessary and sufficient provided that $e^{i \desangle{U}{j}}$ is a
 non-degenerate eigenvalue of $U$.)  Hence, the R.H.S.  of
 Eq.~(\ref{E:min_max_des}) must be greater than or equal to $\desangle{U}{j}$.
 On the other hand, by applying a similar convex hull argument to the
 codimension $j-1$ subspace ${\mathcal H}' = \bigoplus_{k=j}^n
 \desspace{U}{k}$, we know that $\max_{|\psi\rangle \in {\mathcal H}'} \arg
 \langle \psi | U | \psi \rangle = \desangle{U}{j}$.  And the maximum value is
 attained by picking $|\psi\rangle = \desket{U}{j}$.  Hence,
 Eq.~(\ref{E:min_max_des}) is true.

 Lastly, we deduce the second inequality in Eq.~(\ref{E:min_max_des_special})
 by putting $j = 1$ in Eq.~(\ref{E:min_max_des}).  And then, we obtain the
 first inequality in Eq.~(\ref{E:min_max_des_special}) by substituting $U$ by
 $U^{-1}$ into the second inequality.
\end{proof}

 Lemma~\ref{Lemma:min_max} is of interest in its own right for it is analogous
 to the famous minmax principle for Hermitian matrices. (See, for example,
 Theorem~6.1 in Ref.~\cite{eigenvalue_book}.)

 We now give the elementary proofs of Theorems~\ref{Thrm:desangle_bound}
 and~\ref{Thrm:desabsangle_bound}.

\begin{proof}[Elementary proof of Theorem~\ref{Thrm:desangle_bound}]
 We only need to show the validity of Eq.~(\ref{E:UV_des_inequality}) as the
 validity of Eq.~(\ref{E:UV_asc_inequality}) follows directly from it.  This
 is because $\ascangle{U^{-1}}{j} = - \desangle{U}{j}$ for all $n$-dimensional
 unitary matrices $U$ and for $j=1,2,\ldots,n$.

 Since $\desangle{U}{1} + \desangle{V}{1} \leq \pi$ and $\ascangle{U}{1} +
 \ascangle{V}{1} > -\pi$, we have the following three cases to consider.
\par\noindent
 Case~(i): $\desangle{U}{1} - \ascangle{U}{1}, \desangle{V}{1} -
  \ascangle{V}{1} < \pi$;
\par\noindent
 Case~(ii): $\pi \leq \desangle{U}{1} - \ascangle{U}{1} < 2\pi$ and
  $\desangle{V}{1} - \ascangle{V}{1} < \pi$;
\par\noindent
 Case~(iii): $\pi \leq \desangle{V}{1} - \ascangle{V}{1} < 2\pi$ and
  $\desangle{U}{1} - \ascangle{U}{1} < \pi$.

 To prove the validity of Eq.~(\ref{E:UV_des_inequality}) for case~(i), we
 apply Lemma~\ref{Lemma:product_phase} to obtain
 \begin{equation}
  \desangle{U V}{1} = \arg \desbra{U V}{1} U \desket{U V}{1} + \arg
  \desbra{U V}{1} V \desket{U V}{1} .
  \label{E:proof1}
 \end{equation}
 Separately applying Eq.~(\ref{E:min_max_des_special}) in
 Lemma~\ref{Lemma:min_max} to the two terms in the R.H.S. of
 Eq.~(\ref{E:proof1}), we have
 \begin{equation}
  \desangle{U V}{1} \leq \desangle{U}{1} + \desangle{V}{1} .
  \label{E:proof2}
 \end{equation}
 Hence, Eq.~(\ref{E:UV_des_inequality}) is valid for case~(i).  Furthermore,
 the equality holds if and only if $\desket{UV}{1} \in \desallspace{U} \cap
 \desallspace{V}$.  This proves the validity of this theorem for case~(i).

 The validity of cases~(ii) and~(iii) follow that of case~(i).  (For
 simplicity, we only consider the reduction from case~(ii) to case~(i) as the
 reduction from case~(iii) to case~(i) is similar.)  Let $U,V$ be a pair of
 unitary matrices satisfying the conditions of case~(ii).  Then
 $\desangle{U}{1} + \desangle{V}{1} - \ascangle{U}{1} - \ascangle{V}{1} <
 2\pi$.  So, we can pick a number $a$ from the non-empty open interval
 \begin{equation}
  a \in \left( \frac{\desangle{U}{1} - \ascangle{U}{1} - \pi}{\desangle{U}{1} -
  \ascangle{U}{1}}, \frac{\pi - \desangle{V}{1} +
  \ascangle{V}{1}}{\desangle{U}{1} - \ascangle{U}{1}} \right) .
  \label{E:a_def}
 \end{equation}
 It is easy to check that $a\in (0,1)$ and that $0 < a \left[ \desangle{U}{1} -
 \ascangle{U}{1} \right]$, $(1-a) \left[ \desangle{U}{1} - \ascangle{U}{1}
 \right]$, $a \left[ \desangle{U}{1} - \ascangle{U}{1} \right] +
 \desangle{V}{1} - \ascangle{V}{1} < \pi$.  As a result, the pair of matrices
 $U^a$ and $V$ satisfies the conditions of this theorem for case~(i) where the
 notation $U^a$ denotes the unitary matrix $\sum_j e^{i a\desangle{U}{j}}
 \desket{U}{j}\desbra{U}{j}$.  Therefore, $\desangle{U^a V}{1} \leq
 \desangle{U^a}{1} + \desangle{V}{1} = a \desangle{U}{1} + \desangle{V}{1}$.
 Further notice that the pair of matrices $U^{1-a}$ and $U^a V$ also obeys the
 conditions of this theorem for case~(i).  Hence, $\desangle{U V}{1} =
 \desangle{U^{1-a} (U^a V)}{1} \leq \desangle{U^{1-a}}{1} + \desangle{U^a V}{1}
 \leq (1-a) \desangle{U}{1} + a \desangle{U}{1} + \desangle{V}{1} =
 \desangle{U}{1} + \desangle{V}{1}$.  Clearly, for case~(ii),
 Eq.~(\ref{E:UV_des_inequality}) becomes an equality if and only if
 $\desket{UV}{1} \in \desallspace{U^{1-a}} \cap \desallspace{U^a V} \cap
 \desallspace{U^a} \cap \desallspace{V} = \desallspace{U} \cap
 \desallspace{V}$.  This proves the validity of this theorem for case~(ii).
\end{proof}

\begin{proof}[Elementary proof of Theorem~\ref{Thrm:desabsangle_bound}] We may
 assume that $\desabsangle{U}{1} + \desabsangle{V}{1} < \pi$ for the theorem is
 trivially true otherwise.  Then, from Eqs.~(\ref{E:UV_des_inequality})
 and~(\ref{E:UV_asc_inequality}) in Theorem~\ref{Thrm:desangle_bound}, we have
 \begin{eqnarray}
  \desabsangle{U V}{1} & = & \max \left[ \desangle{U V}{1} , -\ascangle{U V}{1}
   \right] \nonumber \\
  & \leq & \max \left[ \desangle{U}{1} + \desangle{V}{1} , -\ascangle{U}{1}
   -\ascangle{V}{1} \right] \nonumber \\
  & \leq & \desabsangle{U}{1} + \desabsangle{V}{1} .
  \label{E:proofa}
 \end{eqnarray}

 Suppose $\desangle{U}{1} + \desangle{V}{1} > - \ascangle{U}{1} -
 \ascangle{V}{1}$, then the last inequality in the above equation is an
 equality if and only if $\desangle{U}{1} = \desabsangle{U}{1}$ and
 $\desangle{V}{1} = \desabsangle{V}{1}$.  By the same argument, in the case of
 $\desangle{U}{1} + \desangle{V}{1} < -\ascangle{U}{1} - \ascangle{V}{1}$, the
 last inequality in the above equation is an equality if and only if
 $\ascangle{U}{1} = -\desabsangle{U}{1}$ and $\ascangle{V}{1} =
 -\desabsangle{V}{1}$.  Applying Lemma~\ref{Thrm:desangle_bound} to analyze the
 condition for equality of the first inequality in Eq.~(\ref{E:proofa}), we get
 the necessary and sufficient conditions for equality as stated in this theorem
 for the case of $\desabsangle{U}{1} + \desabsangle{V}{1} < \pi$.  Whereas in
 the case of $\desabsangle{U}{1} + \desabsangle{V}{1} = \pi$, we use a similar
 trick in our elementary proof of Theorem~\ref{Thrm:desangle_bound} by choosing
 a real number $a \in (0,1)$ such that $\desabsangle{U^a}{1},
 \desabsangle{U^{1-a}}{1}, \desabsangle{V}{1}, \desabsangle{U^a}{1} +
 \desabsangle{V}{1} < \pi/2$.  Then, by analyzing the conditions for equality
 for Theorem~\ref{Thrm:desabsangle_bound} for the pairs of unitary matrices
 $U^a$ and $V$, we conclude that the necessary and sufficient conditions stated
 in this theorem is true for the case of $\desabsangle{U}{1} +
 \desabsangle{V}{1} = \pi$.
\end{proof}

 After simple modifications both in the theorems and our proofs, we find the
 infinite-dimensional analogs of Theorems~\ref{Thrm:desangle_bound}
 and~\ref{Thrm:desabsangle_bound}.  Note that $\desangle{U}{j}$'s and the likes
 are no longer well-defined for an infinite-dimensional unitary operator $U$.
 Nevertheless, we can still talk about $\sup\arg (U)$ the supremum of the
 arguments of the spectrum of $U$.  The symbols $\inf\arg (U)$ and $\sup
 \left|\arg\right| (U)$ can be similarly defined.  We now state the extensions
 of Theorems~\ref{Thrm:desangle_bound} and~\ref{Thrm:desabsangle_bound} below.

\begin{theorem}
 Let $U, V$ be two unitary operators acting on the same complex Hilbert space
 with $\sup\arg(U) + \sup\arg(V) \leq \pi$ and $\inf\arg(U) + \inf\arg(V) >
 -\pi$.  Then,
 \begin{subequations}
 \begin{equation}
  \sup\arg (U V) \leq \sup\arg (U) + \sup\arg (V)
  \label{E:extended_des_inequality}
 \end{equation}
 and
 \begin{equation}
  \inf\arg (U V) \geq \inf\arg (U) + \inf\arg (V) .
  \label{E:extended_asc_inequality}
 \end{equation}
 \end{subequations}
 Moreover, the equality of Eq.~(\ref{E:extended_des_inequality}) holds if and
 only if there exists a sequence of eigenkets $\{ |\psi_j\rangle
 \}_{j=1}^\infty$ of $U V$ such that $\lim_{j\rightarrow\infty} \arg \langle
 \psi_j | U V | \psi_j \rangle = \sup\arg (U V), \lim_{j\rightarrow\infty} \arg
 \langle \psi_j | U | \psi_j \rangle = \sup\arg (U)$ and $\lim_{j\rightarrow
 \infty} \arg \langle \psi_j | V | \psi_j \rangle = \sup\arg (V)$.  In a
 similar fashion, the equality of Eq.~(\ref{E:extended_asc_inequality}) holds
 if and only if there exists a sequence of eigenkets $\{ |\psi_j\rangle
 \}_{j=1}^\infty$ of $U V$ such that $\lim_{j\rightarrow\infty} \arg \langle
 \psi_j | U V | \psi_j \rangle = \inf\arg (U V), \lim_{j\rightarrow\infty} \arg
 \langle \psi_j | U | \psi_j \rangle = \inf\arg (U)$ and $\lim_{j\rightarrow
 \infty} \arg \langle \psi_j | V | \psi_j \rangle = \inf\arg (V)$.
 \label{Thrm:extended_desangle_bound}
\end{theorem}

\begin{theorem}
 Let $U, V$ be two unitary operators acting on the same complex Hilbert space.
 Then,
 \begin{equation}
  \sup\left|\arg\right| (U V) \leq \sup\left|\arg\right| (U) + \sup\left|\arg
  \right| (V) .
  \label{E:extended_desabs_inequality}
 \end{equation}
 Moreover, the equality holds if and only if
 \begin{enumerate}
  \item $\sup\left|\arg\right| (U) + \sup\left|\arg\right| (V) \leq \pi$;
  \item there exist a sequence of eigenkets $\{ |\psi_j\rangle \}_{j=1}^\infty$
   of $U V$ such that $\lim_{j\rightarrow\infty} \left| \arg \langle \psi_j | U
   V | \psi_j \rangle \right|$ $= \sup\left|\arg\right| (U V)$; and
  \item
   \begin{enumerate}
    \item $\lim_{j\rightarrow\infty} \arg \langle \psi_j | U | \psi_j \rangle =
     \sup\arg (U) = \sup\left|\arg\right| (U)$ and $\lim_{j\rightarrow\infty}
     \arg \langle \psi_j | V | \psi_j \rangle = \sup\arg (V) = \sup\left|\arg
     \right| (V)$;
     or
    \item $\lim_{j\rightarrow\infty} \arg \langle \psi_j | U | \psi_j \rangle =
     \inf\arg (U) = -\sup\left|\arg\right| (U)$ and $\lim_{j\rightarrow\infty}
     \arg \langle \psi_j | V | \psi_j \rangle = \inf\arg (V) = -\sup\left|\arg
     \right| (V)$.
   \end{enumerate}
 \end{enumerate}
 \label{Thrm:extended_desabsangle_bound}
\end{theorem}

\begin{proof}[Outline proofs of Theorems~\ref{Thrm:extended_desangle_bound}
 and~\ref{Thrm:extended_desabsangle_bound}]
 We can use the convex hull argument in Lemmas~\ref{Lemma:product_phase}
 and~\ref{Lemma:min_max} to show that (1)~$\sup\arg (U V) = \sup\arg
 \langle\phi| U |\phi\rangle + \sup\arg \langle\phi| V |\phi\rangle$ where the
 supremum is taken over all eigenkets $|\phi\rangle$ of $U V$; and
 (2)~$\inf\arg (U) \leq \arg \langle\psi| U |\psi\rangle \leq \sup\arg (U)$ for
 all $|\psi\rangle$ whenever $\sup\arg (U) - \inf\arg (U) < \pi$.  Hence,
 Eq.~(\ref{E:extended_des_inequality}) in
 Theorem~\ref{Thrm:extended_desangle_bound} holds in the case of $\sup\arg (U)
 - \inf\arg (U), \sup\arg (V) - \inf\arg (V) < \pi$.  Furthermore, by examining
 the condition for $\arg \langle\psi| U |\psi\rangle = \sup\arg (U)$ in the
 case of $\sup\arg (U) - \inf\arg (U) < \pi$, it is straight-forward to verify
 the validity of the necessary and sufficient conditions for equality of
 Eq.~(\ref{E:extended_des_inequality}) in the case of $\sup\arg (U) - \inf\arg
 (U), \sup\arg (V) - \inf\arg (V) < \pi$.  Now, we can follow the arguments in
 the proofs of the remaining cases in Theorem~\ref{Thrm:desangle_bound} as well
 as the proof of Theorem~\ref{Thrm:desabsangle_bound} to show the validity of
 Theorems~\ref{Thrm:extended_desangle_bound}
 and~\ref{Thrm:extended_desabsangle_bound}.
\end{proof}

\begin{acknowledgments}
 We thank F.~K. Chow, C.-H.~F. Fung and K.~Y. Lee for their enlightening
 discussions.
 This work is supported by the RGC grant number HKU~700709P of the HKSAR
 Government.
\end{acknowledgments}

\bibliography{qc48.3}
\end{document}